\documentclass{ws-procs961x669} 

\usepackage{hyperref}
 
\begin{document}

\title{On the evolution of inhomogeneous perturbations \\
in the $\Lambda$CDM model and $f\left(R\right)$ modified gravity theories}

\author{T. Schiavone$^*$}

\address{Department of Physics ``E. Fermi'', University of Pisa, Polo Fibonacci,\\
Largo B. Pontecorvo 3, I-56127, Pisa, Italy\\
and\\
INFN, Istituto Nazionale di Fisica Nucleare, Sezione di Pisa, Polo Fibonacci,\\
Largo B. Pontecorvo 3, I-56127, Pisa, Italy\\
$^*$E-mail: tiziano.schiavone@pi.infn.it\\
}

\author{G. Montani}

\address{ENEA, Fusion and Nuclear Safety Department, C.R. Frascati,\\
Via E. Fermi 45, I-00044 Frascati (Rome), Italy\\
and\\
Physics Department, ``Sapienza'' University of Rome,\\
P.le Aldo Moro 5, I-00185 Rome, Italy
}

\begin{abstract}
We focus on weak inhomogeneous models of the Universe at low redshifts, described by the 
Lema\^itre-Tolman-Bondi (LTB) metric. The principal aim of this work is to compare the evolution of inhomogeneous perturbations in the $\Lambda$CDM cosmological model and $f(R)$ modified gravity theories, considering a flat Friedmann-Lema\^itre-Robertson-Walker (FLRW) metric for the background. More specifically, we adopt the equivalent scalar-tensor formalism in the Jordan frame, in which the extra degree of freedom of the $f(R)$ function is converted into a non-minimally coupled scalar field. We investigate the evolution of local inhomogeneities in time and space separately, following a linear perturbation approach. Then, we obtain spherically symmetric solutions in both cosmological models. Our results allow us to distinguish between the presence of a cosmological constant and modified gravity scenarios, since a peculiar Yukawa-like solution for radial perturbations occurs in the Jordan frame. Furthermore, the radial profile of perturbations does not depend on a particular choice of the $f(R)$ function, hence our results are valid for any $f(R)$ model.
\end{abstract}

\keywords{Cosmology, Modified gravity, Inhomogeneous Universe, Late Universe, Large Scale Structure}

\bodymatter

\section*{Introduction}
\label{sec:intro}

The well-known $\Lambda$CDM model \cite{weinberg,peebles}, which
includes a cosmological constant $\Lambda$ and a cold dark matter
(CDM) component, is based on General Relativity (GR), and it provides
a robust and accurate description of our Universe, supported by cosmological
data. Moreover, the Universe appears homogeneous and isotropic at
very large scales \cite{Planck2020} (cosmological principle), and
this kind of geometry is described by the FLRW line element. In this work, based on a paper in preparation \cite{inPreparation},
we investigate possible deviations from two pillars of the standard
$\Lambda$CDM model, focusing on a local inhomogeneous description
of the Universe and modified gravity models. 

Here, we adopt the LTB metric \cite{peebles},
a spherically symmetric solution of the Einstein field equations,
to account for local deviations of the Universe from the homogeneity.
Hence, we regard the weakly inhomogeneous LTB solution as the standard
FLRW geometry plus small spherically symmetric perturbations.

Furthermore, in recent years increasing interest raised to provide
features able to distinguish between GR and modified gravity approaches.
Some open problems in cosmology, such as the Hubble constant tension
\cite{reviewH0}, may require new physics \cite{H0(z)1}. The aim
of the present work is to compare the evolution of linear inhomogeneous
perturbations in the $\Lambda$CDM and $f\left(R\right)$ modified
gravity models \cite{odintsov-f(R),sotiriou}. More in detail,
we used the equivalent formalism in the Jordan frame \cite{odintsov-f(R),sotiriou,book-capozz-faraoni},
and we adopt the $f\left(R\right)$ Hu-Sawicki model \cite{hu-sawicki},
which is a promising theoretical framework to reproduce the cosmic
acceleration in the late Universe via the dynamics of a non-minimally
coupled scalar field without a cosmological constant.

Then, we investigate the dynamics of inhomogeneous perturbations in
the two cosmological models abovementioned. Moreover, the separable
variables method is implemented in the analysis at the first-order
of perturbation, and our approach is the same used in Ref.~\citenum{marcoccia}.
We obtain an analytic expression only for the radial dependence of
linear perturbations, while their time evolution must be numerically
evaluated. The most relevant issue, emerging from our analysis, is
the different morphology of the radial dependence of the LTB solution
in the $\Lambda$CDM formulation and Hu-Sawicki dynamics. Indeed,
a distinctive Yukawa-like behavior occurs in the Jordan frame gravity
regardless of the choice of a specific $f\left(R\right)$ model. Hence,
we have found a peculiar feature to distinguish a modified gravity
model from a cosmological constant scenario.

This work is organized as follows: in Sec.~\ref{sec:modified-gravity}
we introduce the framework of the $f\left(R\right)$ extended gravity
models; in Sec.~\ref{sec: LTB} we adopt the LTB metric; then, in
Sec.~\ref{sec:Evolution-of-inhomogeneous} we implement a perturbation
approach to deal with local inhomogeneities. Finally, we summarize
our work in Sec.~\ref{sec:Conclusion-and-future}. We adopt the metric
signature $\left(-,+,+,+\right)$, and we set $c=1$.

\section{$f\left(R\right)$ models in the Jordan frame}

\label{sec:modified-gravity}

A generalization of the Einstein-Hilbert theory is provided by the
so-called $f\left(R\right)$ modified gravity models \cite{odintsov-f(R),sotiriou}.
In this context, the gravitational Lagrangian density is expressed
as a free function $f$ of the Ricci scalar $R$, namely an extra
degree of freedom compared to GR. The extended field equations in
$f\left(R\right)$ gravity are fourth-order partial differential equations
in the metric tensor components. In particular, if $f\left(R\right)=R$,
the gravitational field equations end up in the Einstein equations
in GR, trivially. 

It is often convenient to reformulate $f\left(R\right)$ gravity in
the scalar-tensor representation in the Jordan frame \cite{odintsov-f(R),sotiriou,book-capozz-faraoni}.
In doing so, the extra degree of freedom of the $f\left(R\right)$
function is converted into a non-minimally coupled scalar field. It
can be checked that the following action
\begin{equation}
S_{J}=\frac{1}{2\,\chi}\,\int_{\Omega}d^{4}x\,\sqrt{-g}\,\left[\phi\,R-V\left(\phi\right)\right]+S_{M}\left(g_{\mu\nu},\psi\right)\label{eq: azione jordan frame}
\end{equation}
is dynamically equivalent to the action in the $f\left(R\right)$
metric formalism, if $f^{\prime\prime}\left(R\right)\neq0$, where
we have defined a scalar field $\phi=f^{\prime}\left(R\right)=df/dR,$
and a scalar field potential $V\left(\phi\right)=\phi\,R\left(\phi\right)-f\left(R\left(\phi\right)\right)$.
In Eq. \eqref{eq: azione jordan frame}, $\chi\equiv8\,\pi\,G$ is
the Einstein constant, $G$ is the Newton constant, $g$ is the determinant
of the metric tensor $g_{\mu\nu}$, while $S_{M}$ is the matter action,
and $\psi$ denotes matter fields. The advantage of the Jordan frame
is that the corresponding field equations \cite{sotiriou}
are now second-order differential equations, but one has to deal with
the non-minimally coupling. 

The increasing attention to these extended gravity models is motivated
by the consideration that the actual cosmic acceleration of the Universe
might be obtained by geometry rather than a cosmological constant. A
geometrical modification, indeed, could be regarded as an effective
matter source. 

Among several suggested $f\left(R\right)$ models \cite{odintsov-f(R),sotiriou,starob,tsujik},
we focus on the Hu-Sawicki proposal \cite{hu-sawicki}, which is
largely studied in the late Universe. The deviation $F\left(R\right)\equiv f\left(R\right)-R$
with respect to the GR scenario for the Hu-Sawicki (HS) model with
$n=1$ assumes the following form
\begin{equation}
F\left(R\right)=-m^{2}\frac{c_{1}\,R/m^{2}}{c_{2}\,R/m^{2}+1}\,,\label{eq:F(R)-HS}
\end{equation}
where $m^{2}\equiv\frac{\chi\,\rho_{m0}}{3}$ is related to the present
matter density $\rho_{m0}$, while $c_{1}$ and $c_{2}$ are dimensionless
parameters. It can be checked \cite{hu-sawicki} that an effective cosmological constant
is obtained for $R\gg m^{2}$.
Furthermore, $c_{1}$ and $c_{2}$ can be constrained
\cite{hu-sawicki} specifying $F_{R0}\equiv dF\,/\,dR\,\left(z=0\right)$ at the present cosmic time (redshift $z=0$).

Finally, we write the scalar field potential $V\left(\phi\right)$
in the Jordan frame for the HS model with $n=1$:
\begin{equation}
V\left(\phi\right)=\frac{m^{2}}{c_{2}}\,\left[c_{1}+1-\phi-2\sqrt{c_{1}\,\left(1-\phi\right)}\right]\,.\label{eq:Hu-Sawicki-potential}
\end{equation}
It can be shown \cite{potenzialeHS} that the potential \eqref{eq:Hu-Sawicki-potential}
ends in an asymptotically stable de Sitter Universe.

\section{Inhomogeneous solutions in GR and in the Jordan frame}

\label{sec: LTB}

An inhomogeneous Universe can be described by using the LTB
line element \cite{peebles}, which is a spherically symmetric solution
of the Einstein equations. The Universe in the LTB geometry appears
inhomogeneous, but isotropic, from a single preferred point located
at the center of the spherical symmetry. For instance, a cosmological
dust or a spherical overdensity mass are well described by such a
metric.

In the synchronous gauge, the LTB line element may be regarded as
a generalization of the FLRW one, and it is given by
\begin{equation}
ds^{2}=-dt^{2}+e^{2\alpha}dr^{2}+e^{2\beta}\,\left(d\theta^{2}+sin^{2}\theta\,d\phi^{2}\right)\,,\label{eq:LTBmetric}
\end{equation}
where $r$ is the radial distance from the preferred point, while
$\alpha=\alpha\left(t,\,r\right)$ and $\beta=\beta\left(t,\,r\right)$
are two metric functions. 

Considering a pressure-less dust ($p=0$) and a cosmological constant $\Lambda$ in GR, we can write the Einstein equations in
the LTB metric. For instance, the 01 component of field equations
is written as 
\begin{equation}
\frac{\dot{\beta}^{\prime}}{\beta^{\prime}}-\dot{\alpha}+\dot{\beta}=0\label{eq:LTB-01-GR}
\end{equation}
where $\dot{\left(\right)}$ and $\left(\right){}^{\prime}$ denote
derivatives with respect to time $t$ and radial coordinate $r$,
respectively. The other non-zero equations are the 00 and 11 components \cite{inPreparation}, while different components vanish due to spherical symmetry. 

It is possible to simplify the LTB metric within the framework of
GR, using Eq. \eqref{eq:LTB-01-GR}, which allows us to find a
relation between the metric functions $\alpha$ and $\beta$. Following
the approach in Ref.~\citenum{peebles}, the LTB line element \eqref{eq:LTBmetric}
becomes
\begin{equation}
ds^{2}=-dt^{2}+\frac{\left[\left(a\,r\right){}^{\prime}\right]{}^{2}}{1-r^{2}\,K^{2}}dr^{2}+\left(a\,r\right){}^{2}\,\left(d\theta^{2}+sin^{2}\theta\,d\phi^{2}\right)\,,\label{eq:LTBmetric-GR-simpler}
\end{equation}
where $a=a\left(t,r\right)=e^{\beta}/r$ is a generalized
scale factor in a inhomogeneous geometry, and $K=K\left(r\right)$.
Note that if $a\left(t,r\right)$ and $K\left(r\right)$ are independent
of $r$, the metric \eqref{eq:LTBmetric-GR-simpler} coincides with
the FLRW line element, describing an isotropic and homogeneous geometry.
Moreover, the 00 and 11 components of the Einstein equations in the
LTB metric provides a generalization of the Friedmann equations in
the $\Lambda$CDM model. 

Concerning the $f\left(R\right)$ modified gravity scenario in the
equivalent Jordan frame, the 01 component of the gravitational field
equations in the LTB metric is written as
\begin{equation}
\frac{\dot{\beta}^{\prime}}{\beta^{\prime}}-\dot{\alpha}+\dot{\beta}=-\frac{1}{2\,\phi\,\beta^{\prime}}\,\left(\dot{\phi}^{\prime}-\dot{\alpha}\,\phi^{\prime}\right)\,.\label{eq: 01 LTB Jordan frame}
\end{equation}
Furthermore, we can compute \cite{inPreparation} also the 00 and 11 of field equations,
as well as the scalar field equation in the Jordan frame.
Observe that $\phi$, $\rho$, $\alpha$, and $\beta$ are all functions
of $t$ and $r$ in a inhomogeneous scenario.

Comparing Eq. \eqref{eq: 01 LTB Jordan frame} with the respective
equation \eqref{eq:LTB-01-GR} obtained in GR, an extra term occurs
in the Jordan frame due to non-minimal coupling between $\phi$ and
the metric functions $\alpha$, $\beta$. As a consequence, we can
not relate $\alpha$ and $\beta$ to rewrite the LTB metric as in
the GR scenario. Moreover, the scalar field potential $V\left(\phi\right)$
does not affect Eq. \eqref{eq: 01 LTB Jordan frame}, but it is
included in the remaining field equations. $V\left(\phi\right)$ represents
an extra degree of freedom with respect to GR, and the dynamics in
the Jordan frame deviates from that of the Einstein theory.

\section{Perturbation approach in GR and in the Jordan frame}
\label{sec:Evolution-of-inhomogeneous}

In this Section, we investigate the different evolution of local inhomogeneities
within the framework of GR and modified gravity \cite{inPreparation}. Here, we stress one
important difference: we can not adopt the LTB metric in the form
\eqref{eq:LTBmetric-GR-simpler} in the Jordan frame, but we have
to refer to the line element in Eq. \eqref{eq:LTBmetric}, as motivated
in Sec.~\ref{sec: LTB}. 

Then, we consider the LTB metric to describe local inhomogeneities
as small linear perturbations with respect to a flat FLRW metric as
background. Hence, we can write all the physical quantities as a background
term, denoted with a bar, plus a linear perturbation,
which depends on $t$ and $r$, and it is denoted with $\delta$.
For instance, the energy density is written as $\rho\left(t,r\right)=\bar{\rho}\left(t\right)+\delta\rho\left(t,r\right)\,.$
Similarly, we rewrite the scale factor $a\left(t,r\right)$ in GR,
the scalar field $\phi\left(t,r\right)$, the metric functions $\alpha\left(t,r\right)$,
and $\beta\left(t,r\right)$ in the Jordan frame. We also assume that
perturbation terms are small corrections if compared with respective
background quantities, i.e. $\delta\rho\ll\bar{\rho}$. Furthermore,
we consider $K^{2}\left(r\right)$ in the LTB metric \eqref{eq:LTBmetric-GR-simpler}
as a linear term, since it is related to inhomogeneities. It should
be noted that, comparing the LTB line element \eqref{eq:LTBmetric}
and a flat FLRW metric, we have to require two constraints: $\bar{\alpha}\left(t\right)=\ln\left(\bar{a}\left(t\right)\right)$
and $\bar{\beta}\left(t,r\right)=\ln\left(\bar{a}\left(t\right)\,r\right)$.

Concerning the scalar-tensor formalism in the Jordan frame, we expand
the scalar field potential $V\left(\phi\right)$ to the first order
in $\delta\phi$. It can be easily checked that the zeroth order
term $V\left(\bar{\phi}\left(t\right)\right)$ depends only on the
background scalar field \cite{inPreparation}. 

Once we have split background and linear contributions, we focus
on the dynamics in the $\Lambda$CDM cosmological model and in the
Jordan frame gravity. Hence, we rewrite the field equations in GR
and in the Jordan frame, separating background and perturbation terms. 

\subsection{Comparing background solutions}

To study the time evolution, we define a dimensionless variable $\tau=t/t_{0}$,
where $t_{0}$ is the actual time in the synchronous gauge. Note that
$t_{0}$ is approximately $t_{0}\approx1/H_{0}$ in terms of the Hubble
constant. 

In GR, the 00 component of the Einstein equation in the LTB geometry
becomes the Friedmann equation in the FLRW metric at background level.
We set $\bar{a}_{0}=\bar{a}\left(\tau=1\right)$ today, and we neglect
relativistic componenents in the late Universe. Hence, we obtain an
analytical solution \cite{inPreparation}: the evolution of the background scale factor
in terms $\tau$ is written as
\begin{equation}
\bar{a}\left(\tau\right)=\left(\frac{\Omega_{m0}}{\Omega_{\Lambda0}}\right)^{1/3}\,\,\left\{ \sinh\left[\frac{3}{2}\,\sqrt{\Omega_{\Lambda0}}\,\left(\tau-1\right)+\textrm{arcsinh}\left(\sqrt{\frac{\Omega_{\Lambda0}}{\Omega_{m0}}}\right)\right]\right\} ^{2/3}\,.\label{eq:background-scale-factor-GR}
\end{equation}
We used the cosmological density parameters $\ensuremath{\Omega_{m0}=\rho_{m0}/\rho_{c0}}$,
and $\Omega_{\Lambda0}=\Lambda/\,\left(\chi\,\rho_{c0}\right)$ for
matter and cosmological constant components, respectively, where $\ensuremath{\rho_{c0}=3H_{0}^{2}/\chi}$
is the actual critical energy density of the Universe. We recall also
that $\bar{\rho}\left(\tau\right)\propto\bar{a}\left(\tau\right)^{-3}$
for a matter component. 

\begin{figure} 
\centering
\includegraphics[scale=0.148]{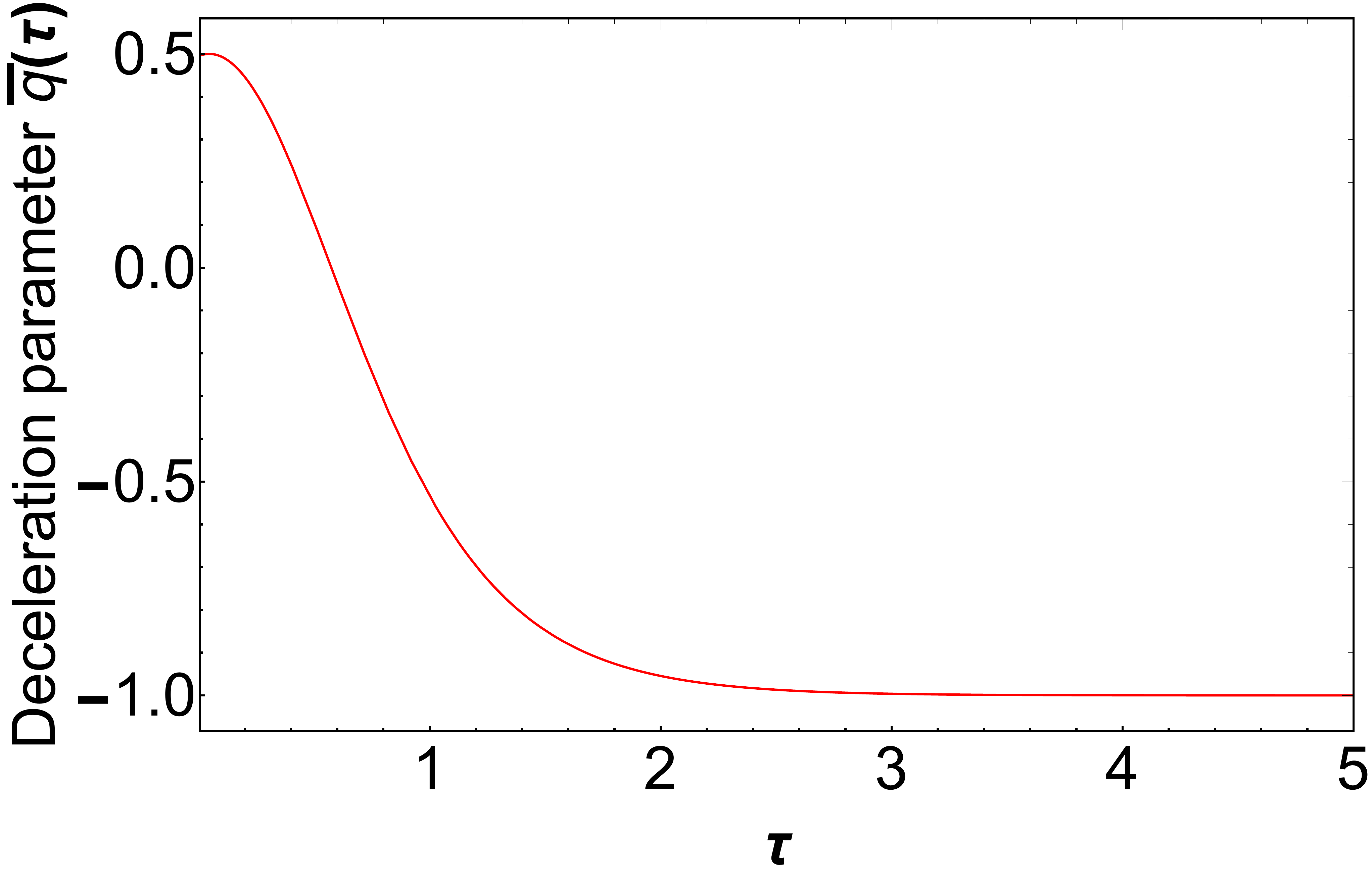} 
\caption{Background deceleration parameter $\bar{q}(\tau)$ in the FLRW geometry in terms of a dimensionless variable $\tau$. The background scenario in the $\Lambda$CDM model are graphically the same comparing it with the one in the $f(R)$ Hu-Sawicki model in the Jordan frame.} \label{figLCDM-background} 
\end{figure}

In Fig.~\ref{figLCDM-background} we plot the deceleration parameter
$\bar{q}\left(\tau\right)\equiv-\ddot{\bar{a}}/\,\left(\bar{a}\,\bar{H}^{2}\right)$,
fixing $\Omega_{m0}=0.3111$ and $\Omega_{\Lambda0}=1-\Omega_{m0}$
from Ref.~\citenum{Planck2020}. Here, the dot denotes the derivative with
respect to $\tau$. The scale factor $\bar{a}\left(\tau\right)$ increases
for growing values of $\tau$, and note that $\bar{q}\rightarrow-1$
in the future. 

We point out that the solution \eqref{eq:background-scale-factor-GR}
in the $\Lambda$CDM cosmological model is viable only at late times
($\tau\gg10^{-6}$) in the Universe, otherwise we have to include
relativistic components, and solve numerically the Friedmann equation. 

Now, we focus on the background field equations in the Jordan frame
gravity. Starting from the 01, 00, 11 components of the gravitational
field equations in the LTB metric and the scalar field equation, if
we do not include local inhomogeneities, we get the respective equation
system in the Jordan frame for the background FLRW metric (see the
field equations in Ref.~\citenum{sotiriou}). We use the Hu-Sawicki model
in the Jordan frame, which is characterized by the scalar field potential
in Eq. \eqref{eq:Hu-Sawicki-potential}. 

\begin{figure} 
\centering
\includegraphics[scale=0.14]{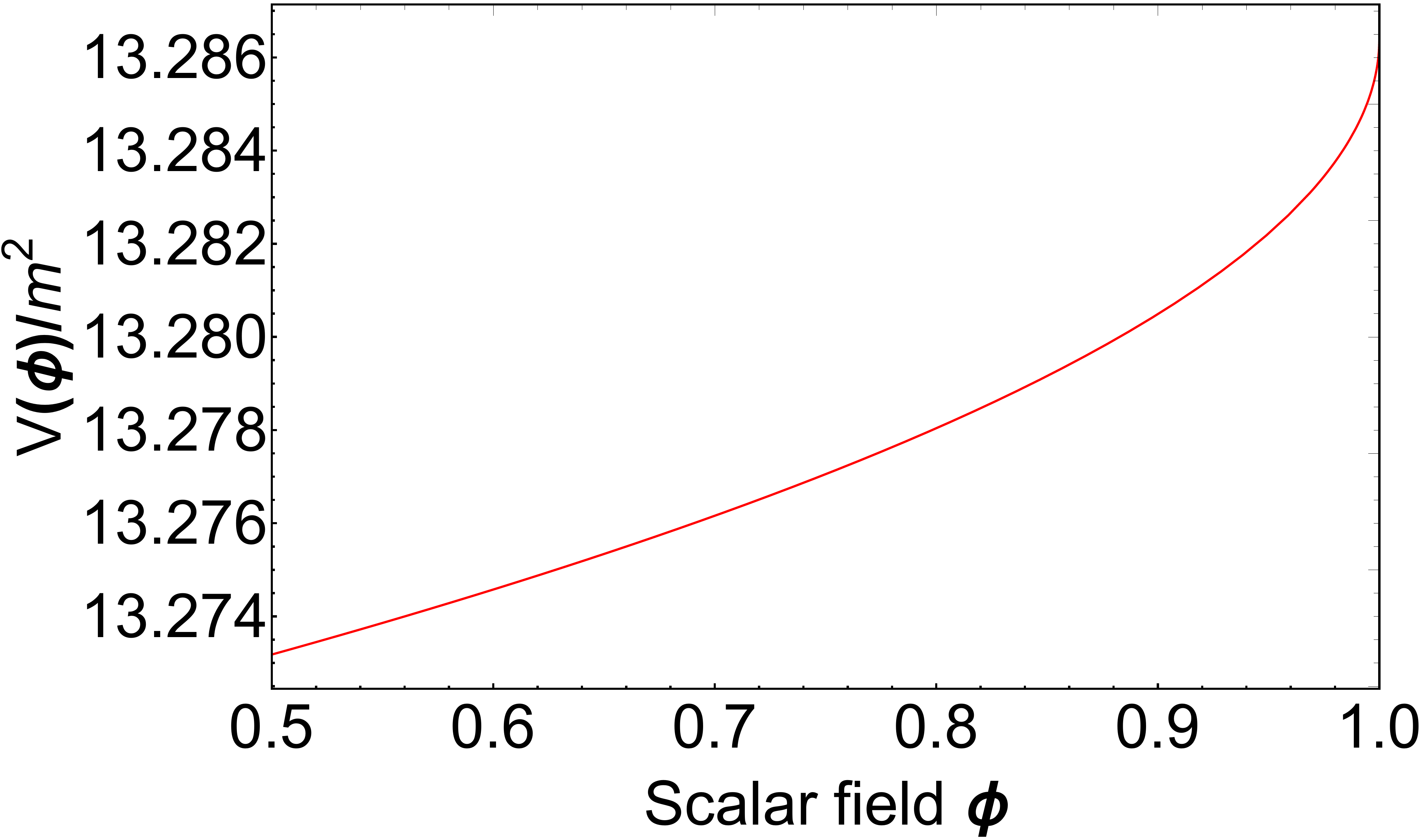}  \quad \includegraphics[scale=0.132]{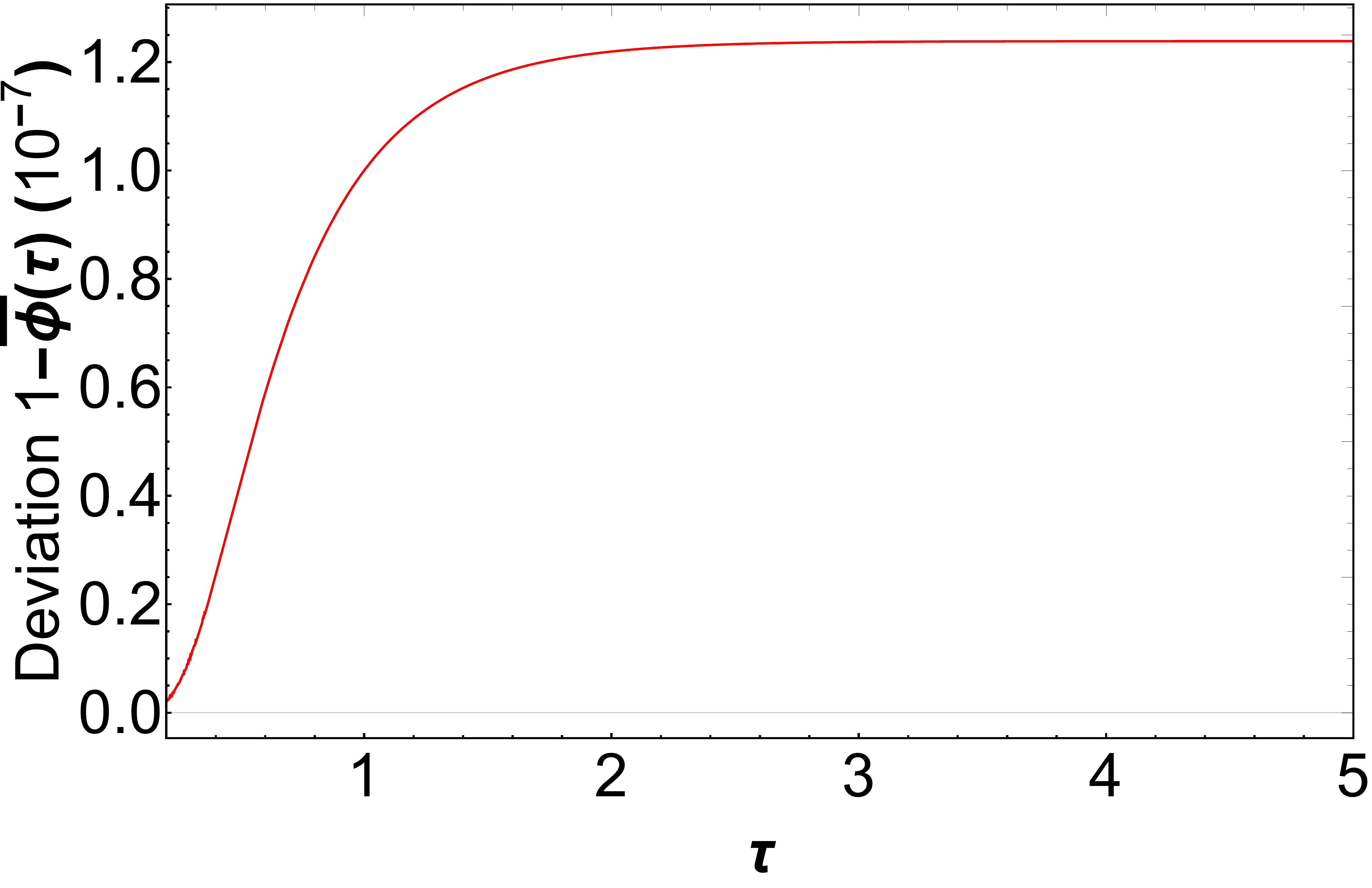} 
\caption{Plots referred to the Hu-Sawicki model in the Jordan frame, adopting the FLRW metric (background solution). Left panel: scalar field potential of the Hu-Sawicki model in the Jordan frame. Right panel: deviation $|1-\bar{\phi}\left(\tau\right)|$ from the GR scenario ($\bar{\phi}=1$) using a scale $10^{-7}$.} \label{figHS-background} 
\end{figure}

To solve numerically the equation system, we fix $\Omega_{m0}$ at
the same value used in the previous study for the $\Lambda$CDM model,
and we choose \cite{hu-sawicki} $\left|F_{R0}\right|=10^{-7}$ to
obtain $c_{1}=2.0\cdot10^{6}$ and $c_{2}=1.5\cdot10^{5}$. We set
the free parameters in a way that the Hu-Sawicki model in the Jordan
frame is very close to the $\Lambda$CDM scenario for the background
level. In Fig.~\ref{figHS-background} note that the scalar field
potential \eqref{eq:Hu-Sawicki-potential} exhibits a slow-roll mimicking
a cosmological constant. 

Finally, we solve numerically the equation system in the Jordan frame \cite{inPreparation},
imposing the conditions at $\tau=1:$ $\bar{\phi}\left(1\right)=1-\left|F_{R0}\right|$
and $d\bar{\phi}/d\tau\left(1\right)=0$. The resulting plot
of the deceleration parameter $\bar{q}\left(\tau\right)$ are graphically
undistinguishable from the respective ones in the $\Lambda$CDM model
(see Fig.~\ref{figLCDM-background}). Concerning the evolution of
the scalar field, we plot in Fig.~\ref{figHS-background} the deviation
$\left|1-\bar{\phi}\left(\tau\right)\right|$ from the GR limit ($\bar{\phi}=1$)
with a scale $10^{-7}$. Note that the dynamics of the Jordan frame
is very close to the one in $\Lambda$CDM model at the background
order for any $\tau$. Now, we can move on to the linearized solutions
to highlight the differences between the two cosmological models.

\subsection{Comparing linear solutions}

To examine separately the time and radial evolution of local inhomogeneities,
we use the separable variables method at the first-order perturbation
theory. So, we factorize all the linear perturbations into time
and radial functions. More specifically, we rewrite the perturbations
of the scale factor $\delta a\left(\tau,r\right)\equiv\delta\texttt{a}\left(\tau\right)\,\delta\mathfrak{a}\left(r\right)$,
the metric functions $\delta\alpha\left(\tau,r\right)\equiv\delta A\left(\tau\right)\,\delta\mathcal{A}\left(r\right)$
and $\delta\beta\left(\tau,r\right)\equiv\delta B\left(\tau\right)\,\delta\mathcal{B}\left(r\right)$,
the energy density $\delta\rho\left(\tau,r\right)\equiv\delta P\left(\tau\right)\,\delta\varrho\left(r\right)$,
and the scalar field $\delta\phi\left(\tau,r\right)\equiv\delta\Phi\left(\tau\right)\,\delta\varphi\left(r\right)$.
The factorization is assumed at the linear level to find analytical
solutions, but the separation of variables is not a general
method, if we also include non-linear terms. 

Then, we investigate the effect of inhomogeneities in the dynamics,
according to the cosmological model considered. We focus on the linearized
equations in the LTB metric within the framework of the $\Lambda$CDM
cosmological model. 

\begin{figure} 
\centering
\includegraphics[scale=0.15]{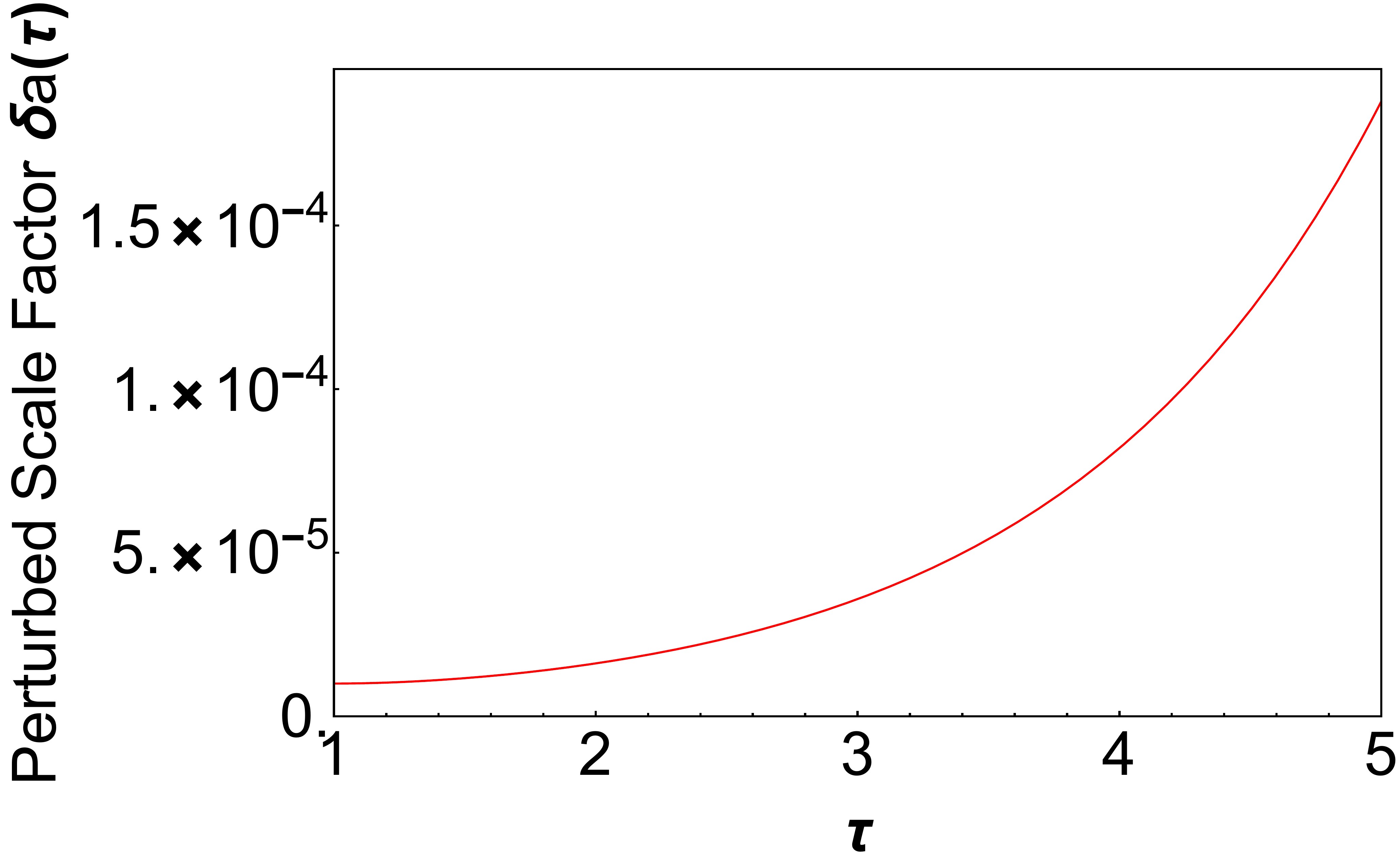}  \quad \includegraphics[scale=0.14]{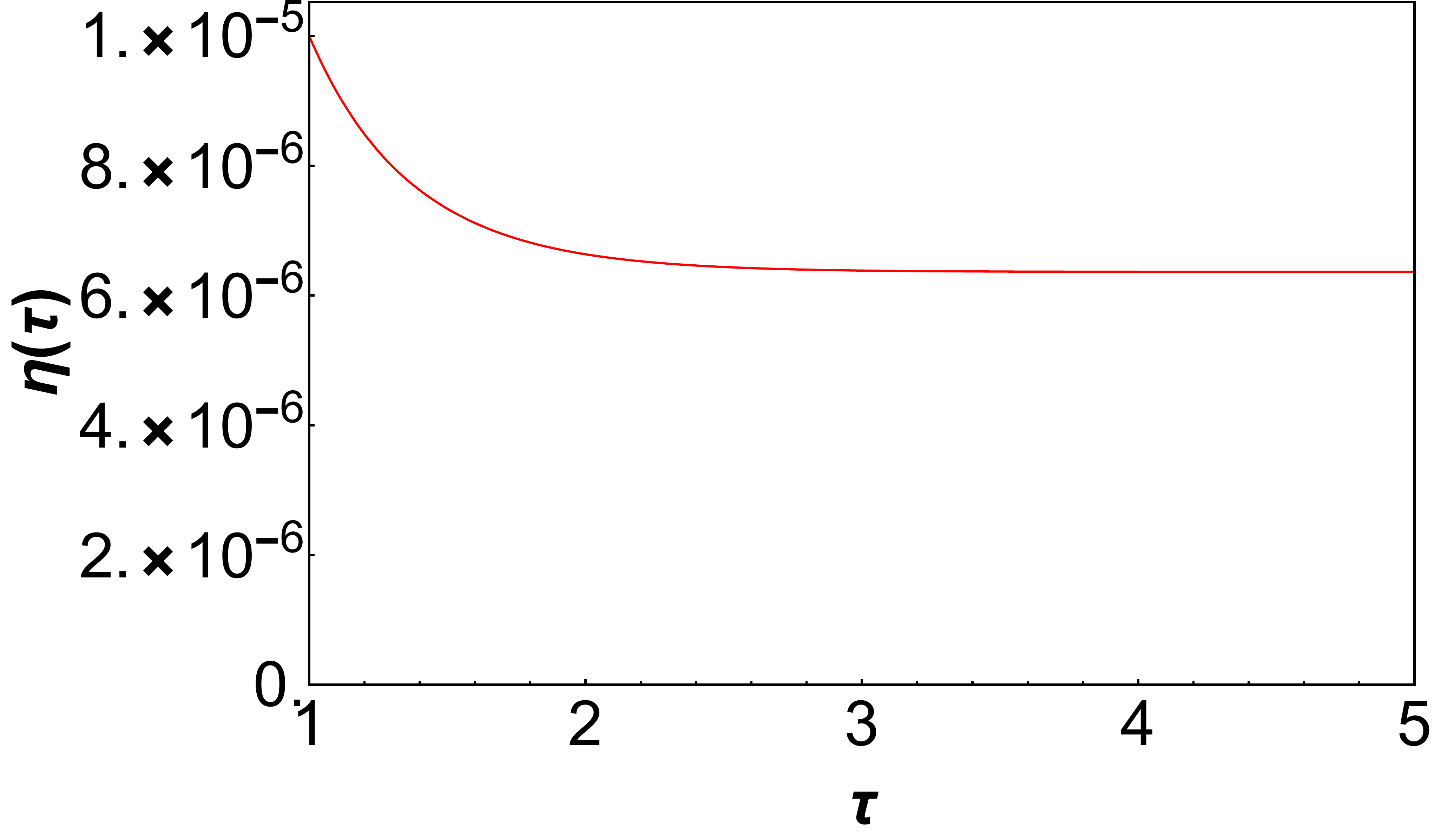}
\caption{Left panel: time evolution of the linear perturbation of the scale factor $\delta\text{a}\left(\tau\right)$. Right panel: the ratio between the linear deviation and background scale factor.} 
\label{figpertGR} 
\end{figure}

Using the factorization abovementioned, it is easy to show that the
11 component of the linearized Einstein equations in the LTB metric
can be separated in two parts \cite{inPreparation}. The radial behavior is $\delta\mathfrak{a}\left(r\right)=K^{2}\left(r\right)\,,$while we have an ordinary differential equation in
terms of $\tau$ for the time evolution: 
\begin{equation}
\delta\ddot{\texttt{a}}\left(\tau\right)+\frac{\dot{\bar{a}}\left(\tau\right)}{\bar{a}\left(\tau\right)}\,\delta\dot{\texttt{a}}\left(\tau\right)-\delta\texttt{a}\left(\tau\right)\,\left[\frac{\ddot{\bar{a}}\left(\tau\right)}{\bar{a}\left(\tau\right)}+\left(\frac{\dot{\bar{a}}\left(\tau\right)}{\bar{a}\left(\tau\right)}\right)^{2}\right]=0\,.
\end{equation}
This equation is solved numerically, once you specify the background
solution \eqref{eq:background-scale-factor-GR}. We impose the following
conditions at $\tau=1$ today: $\delta\texttt{a}\left(1\right)=10^{-5}$
and $\delta\dot{\texttt{a}}\left(1\right)=0$. In Fig.~\ref{figpertGR} we plot the numerical solution, after we have defined
the ratio $\eta\left(\tau\right)$ between the perturbed scale factor
and the background respective quantity, namely $\eta\left(\tau\right)\equiv\left|\delta\texttt{a}\left(\tau\right)\,/\,\bar{a}\left(\tau\right)\right|$.
Note that perturbations of the scale factor remain stable, since
the background scale factor $\bar{a}\left(\tau\right)$ dominates
for any $\tau$ in the late Universe, i.e. $\eta\ll1$. 

After that, we consider the 00 component of the linearized Einstein
equations in the LTB metric, and we combine it with the continuity
equation. After long but straightforward calculations \cite{inPreparation}, we obtain $\delta\mathfrak{a}\propto1/r^{3}$
in GR. According to the cosmological principle, inhomogeneities decay
for increasing values of $r$. Furthermore, we also obtain that the
perturbation of the energy density for a matter component follows
the same dependence of the respective background quantity in the FLRW
metric, namely $\delta P\left(\tau\right)\propto1/\bar{a}^{3}\left(\tau\right)$.
However, we can provide $\delta P\left(\tau\right)\ll\bar{\rho}\left(\tau\right)$
at any times, setting the constant of integration.

Now, we repeat the same approach in the modified gravity dynamics
in the Jordan frame, considering the corresponding linearized field
equations in the LTB metric. Expanding the scalar field potential
and its derivative around the background scalar field, as mentioned
at the beginning of Sec.~\ref{sec:Evolution-of-inhomogeneous}, it
can be proved that the potential $V\left(\phi\right)$ is involved
only in the time evolution of the linearized field equations \cite{inPreparation}. As a
result, we expect that the radial evolution of local inhomogeneities
is completely independent of the choice of a particular $f\left(R\right)$
model, since $V\left(\phi\right)$ is directly related to $f\left(R\right)$. 

In what follows, we do not focus on the time evolution of perturbations,
but we just mention that we find a numerical solution using again
the Hu-Sawicki model (Sec.~\ref{sec:modified-gravity}), and we have
checked that inhomogeneous perturbations remain smaller than respective
background contribution for any time $\tau$. 

To obtain an analytical solution for the radial part with the separable
variables method, we assume two simplifying conditions: $\delta\dot{A}\left(\tau\right)=\delta\dot{B}\left(\tau\right)$
and $\delta\varrho\left(r\right)\propto\delta\varphi\left(r\right)$.
After long but straightforward calculations \cite{inPreparation}, combining the linearized
field equations, we obtain a Yukawa-like solution for the radial part
of the linear perturbation of the scalar field 
\begin{equation}
\delta\varphi\left(r\right)=\frac{C}{r}\,\exp\left(-\frac{r}{r_{c}}\right)\,,
\label{eq:yukawa-scalar-field-1}
\end{equation}
where $C$ and $r_{c}$ are constants. Moreover, $\delta\varrho\left(r\right)$ shows the same evolution, since we recall that it is proportional to $\delta\varphi\left(r\right)$. We also obtain a similar behavior
for $\delta\mathcal{A}\left(r\right)$ and $\delta\mathcal{B}\left(r\right)$.
It should be emphasized that modified gravity introduces a typical
spatial scale, $r_{c}$, such that for $r\gg r_{c}$ the inhomogeneities
decay faster than ones in GR. We stress again that this kind of solutions
applies to any $f(R)$ model, and the radial evolution is different
from the one in GR. 

\section{Summary and conclusions}
\label{sec:Conclusion-and-future}

We have investigated local inhomogeneities in the LTB metric regarded
as small deviations from a flat FLRW background metric. To date, there
are no modified gravity models that predict significant deviations
from GR, reconciling all possible cosmological data. Hence, to try
to discriminate between several cosmological models, it is crucial
to test gravity in different regimes or by using other techniques. Here,
we have suggested one possible method, studying the different dynamics
of inhomogeneous perturbations. Our results have pointed out a distinctive
element in the evolution of local inhomogeneities of the Universe
from a theoretical point of view, allowing to distinguish between
the $\Lambda$CDM cosmological model and $f\left(R\right)$ modified
gravity theories. We have shown that the radial evolution of inhomogeneous
perturbations within the framework of the Jordan frame gravity is
independent of the scalar field potential $V(\phi)$: the distinctive
radial solution is a feature of any $f(R)$ model. Furthermore, we have obtained a Yukawa-like solution for the radial perturbations in the Jordan frame, which is completely different from the one in GR. This work may
be an interesting arena to account for the effects of local inhomogeneities in cosmological observables \cite{fanizza}, when forthcoming missions such as Euclid
Deep Survey \cite{Euclid}, will be able to test the large-scale
properties of the Universe.


\end{document}